\documentclass[twocolumn,twoside,slac]{revtex4}
\usepackage{graphicx}
\usepackage{fancyhdr}
\usepackage{comment}
\begin{document}


\title{Grid Data Management in Action: Experience in Running and
  Supporting Data Management Services in the EU DataGrid Project}

%

\author{Heinz Stockinger, Flavia Donno, Erwin Laure, Shahzad Muzaffar,
Peter Kunszt}
\affiliation{CERN, European Organization for Nuclear Research, CH-1211
  Geneva 23, Switzerland}
\author{Giuseppe Andronico}
\affiliation{INFN Catania, Via S. Sofia 64, I-95123 Catania, Italy}
\author{Paul Millar}
\affiliation{University of Glasgow, Glasgow, G12 8QQ, Scotland}

\begin{abstract}
In the first phase of the EU DataGrid (EDG) project, a Data Management
System has been implemented and provided for deployment. The
components of the current EDG Testbed are: a prototype of a Replica
Manager Service built around the basic services provided by Globus, a
centralised Replica Catalogue to store information about physical
locations of files, and the Grid Data Mirroring Package (GDMP) that is
widely used in various HEP collaborations in Europe and the US for
data mirroring. During this year these services have been refined and
made more robust so that they are fit to be used in a pre-production
environment. Application users have been using this first release of
the Data Management Services for more than a year. In the paper we
present the components and their interaction, our implementation and
experience as well as the feedback received from our user
communities. We have resolved not only issues regarding integration
with other EDG service components but also many of the
interoperability issues with components of our partner projects in
Europe and the U.S. The paper concludes with the basic lessons learned
during this operation. These conclusions provide the motivation for
the architecture of the next generation of Data Management Services
that will be deployed in EDG during 2003.

\end{abstract}

\maketitle

\thispagestyle{fancy}


%
%
\section{\label{introduction}Introduction}

Data management is one of the key features of a Data Grid where large
amounts of data are distributed and/or replicated to remote sites,
potentially all over the world.  In general, a Data Grid needs to
provide features of a pure computational Grid~\cite{Hoschek00}
(resource discovery, sharing etc.) as well as more specialised data
management features like replica management which is the main
focus of this article.

The European DataGrid project~\cite{EDG} (also referred to as \emph{EDG} in
this article), one of the largest Data Grid projects today, has a main
focus on providing and deploying such data replication tools. Although
the project officially started in January 2001, prototype
implementations  started already in early 2000 and a first data
management architecture was presented in~\cite{Hoschek00}. Thus,
within the project there is already a well-established experience in providing
replication tools and deploying them on a large-scale testbed. 

Since interoperability of services and international collaborations on
software development are of major importance for EDG as well
as other Grid projects in Europe, the U.S. etc., the first set of data
management tools (i.e. replication tools) provided and presented here, are
based on established de-facto standards in the Grid community. In
addition, for parts of the software presented here, EDG has
development and deployment collaborations with partner projects like
PPDG~\cite{PPDG}, DataTAG~\cite{datatag} and LCG~\cite{lcg}.

In this article, we present our first set of replication tools that
have been deployed on the European DataGrid testbed. These tools are included
in release 1.4 of the EDG software system and are thus regarded as the
first prototype of the data management software system. Details about the
architecture, software features and experience is given.

The article is organised as follows. In Section~\ref{architecture} we
first  outline briefly the data management challenge and present the
architecture that we established for EDG release 1.4. The replication
tools GDMP (Grid Data Mirroring package) and edg-replica-manager are
presented in the context of the data management architecture. Implementation
details of these tools and a detailed discussion on their differences in
design and usage are presented in Section~\ref{implementation}. Their
deployment in several testbeds and some historical background about
the deployment is given in Section~\ref{deployment}. Since these
replication tools are supposed to be replaced by second generation
replication tools, we briefly introduce them in
Section~\ref{future} since the experience that has been gained
deploying EDG release 1.4 provided vital input for this new
development.

%
%
\section{\label{architecture}Problem Domain, Requirements, Architecture }

In the following section we first describe the data management domain
with its requirements and then a simplified architecture for our first
generation replication tools, i.e. the ones deployed in EDG release
1.4. More details on EDG releases is given in Section~\ref{deployment}.

\subsection{Basic Requirements}
\label{requirements}

In this section we outline the main design features that we have
chosen to meet the requirements of particular data intensive
application domains. We first start with a basic example and then
summarise the basic requirements that are tackled by our replication tools.

\begin{figure}[t,h]
\centering
\includegraphics[width=80mm]{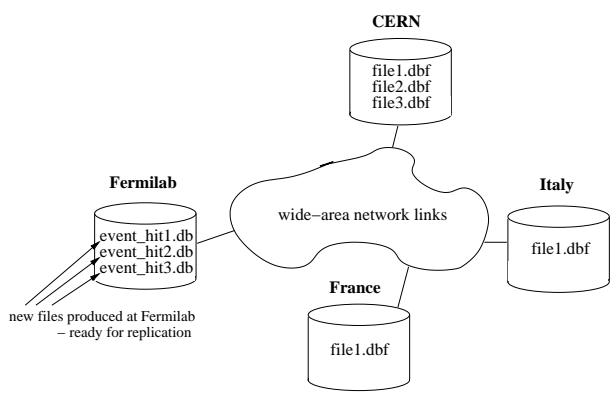}
\caption{Basic replication example.} \label{replication-example}
\end{figure}

In a typical Data Grid, large amounts of data that are stored in
read-only files need to be replicated in a secure and efficient
way~\cite{GDMP1}. As a basic file replication example we consider a Data Grid
that consists of four sites (CERN in Switzerland, Fermilab in the
U.S., Italy and France) as depicted in
Figure~\ref{replication-example}. In the example, new files have been
created at the site ``Fermilab'' and are now ready to be replicated to
remote sites. Several remote sites (e.g. the site CERN) is interested
in having files locally and thus would like to replicate the newly
created files to its local data store. End-users can then access
replicas at both sites and might want to retrieve files with the
lowest access latency.

Note that for simplicity we deal with read-only files and leave the
update-synchronisation problem open for future work as pointed out
in~\cite{Duellmann2001}.


In the following list the major data management and replication
requirements are summarised and the need for a specific software
solution is outlined. Most of these solutions are covered by our
replication tools discussed in the article.

\begin{itemize}

\item Files need to be transferred (copied) between several large data
  stores that reside at distributed sites. - \emph{need for secure and
    efficient file transfer mechanism} (e.g. GridFTP or equivalent)

\item Since replication implies that identical file copies exist, 
  replicas need to be uniquely identified through logical and physical
  filenames. - \emph{need for Replica Catalogue for naming and
    locating replicas}

\item Combine file transfer with file cataloguing and present it as an
  atomic transaction to the user. - \emph{need for replica
  management service}

\item Large data stores use secondary and possibly tertiary storage
  devices, i.e. disk and tape systems, respectively. - \emph{need for
  interaction between replica management service and storage service},
i.e. mass storage interface
\end{itemize}

More detailed requirements of Data Grids (in particular data intensive
scientific domains like High Energy Physics) and the data distribution
problem are presented elsewhere~\cite{Hoschek00, GDMP1}.

\subsection{Basic Terminology}

In the remainder of this article, we use the following EDG
terminology:

\begin{itemize}

\item \emph{Storage Element (SE)}: a Storage Element is a data store that
  provides secondary and/or tertiary storage devices as well as a
  data transfer mechanism that allows for file transfers between
  several Storage Elements connected via wide-area network links

\item \emph{Computing Element (CE)}: a computing Element can be regarded as
  a gateway to several Worker Nodes (WN) that are responsible for the execution
  of a user job. It is important to note that data that has been
  produced on Worker Nodes needs to be stored on Storage Elements in
  order to be accessible for subsequent user jobs.

\item \emph{User Interface (UI)}: a User Interface node is a machine where
  application users can log on and have access to the EDG software
  tools. In principle, the UI contains client software tools.

\item \emph{Logical File Name (LFN)}: an LFN uniquely identifies a
  set of identical replicas.

\item \emph{Physical File Name (PFN)}: identifies one file (replica) of a set
  of identical replicas. Note that the terminology changes from time
  to time but it is important to note that the PFN identifies a real
  data file in a Storage Element or a Computing Element (Worker Node). 

\end{itemize}

\subsection{Architecture of the Data Management Services}

\begin{figure}[t,h]
\centering
\includegraphics[width=80mm]{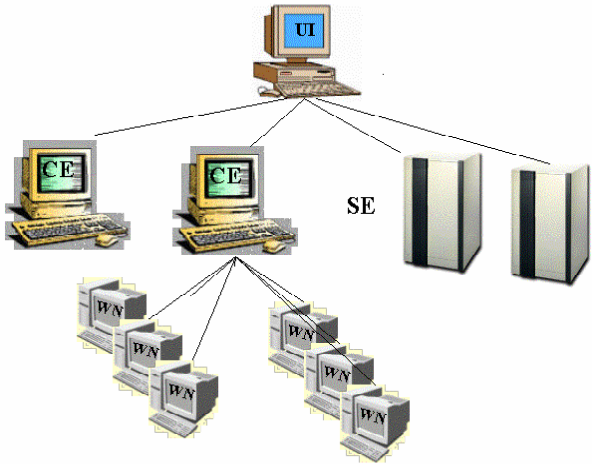}
\caption{Basic topology of EDG testbed (simplified for
  replication tools).} \label{edg-nodes}
\end{figure}

The architecture of our replication tools is based on a typical topology
of Storage Elements, Computing Elements (i.e. Worker Nodes) and User
Interface nodes (UI) as outlined below. This topology is also realised
in the EDG testbed as illustrated in Figure~\ref{edg-nodes}.

File transfer is mainly done between Worker Nodes (of Computing
Elements) and Storage Elements where files are permanently kept and
registered with a Replica Catalogue (RC). In order to introduce new
files to the Grid, they can also be transferred from a UI node to a
Worker Node or directly to the Storage Element. All these use cases
have the following in common: a client-server architecture is required
where client tools have to be available on the User Interface as well
as on Worker Nodes. Server software is mainly required on the Storage
Elements.

In order to meet the requirements outlined in
Section~\ref{requirements}, we have developed replica management tools
in two steps:

\begin{itemize}

\item {\bf GDMP (Grid Data Mirroring Package)}~\cite{GDMP1, GDMP2} for
  replication (mirroring) of file sets between several Storage
  Elements. This was the first replication tool that was developed in
  collaboration between the European DataGrid project and the Particle
  Physics Data Grid (PPDG)~\cite{PPDG} (refer to
  Section~\ref{deployment} for more details on release dates). This
  replication tool also provides a simple interface to Mass Storage
  Systems (MSS).

\begin{figure}[t,h]
\centering
\includegraphics[width=40mm]{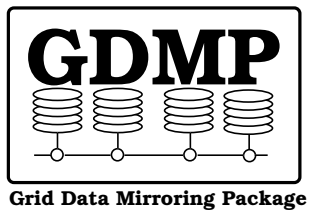}
\caption{Grid Data Mirroring Package (GDMP) - logo.} \label{gdmp-logo}
\end{figure}

\item {\bf edg-replica-manager} was developed in the second year of
  the DataGrid project. It provides some added replication
  functionality that meets additional user requires that were identified during
  the deployment of GDMP in the EDG testbed. In this way, both tools
  complement each other and provide the basic replication
  functionality of the first generation replication tools.

\begin{figure}[t,h]
\centering
\includegraphics[width=40mm]{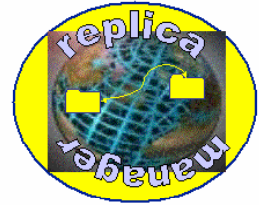}
\caption{edg-replica-manager - logo.} \label{rm-logo}
\end{figure}

\end{itemize}

Both, GDMP and edg-replica-manager are part of the current (as of
March 2003) EDG software release 1.4. 

Both, GDMP and edg-replica-manager use components of the Globus
Toolkit 2 (TM) and thus are based on the current de-facto standard in
Grid computing. Although GDMP and edg-replica-manager are
architecturally different (client-server architecture versus client
side tool only - details are given in Section~\ref{implementation}) - they
have the following architectural components in common:

\begin{itemize}

\item GridFTP~\cite{Allcock2001} for efficient and secure file
  transfer.

\item Grid Security Infrastructure (GSI) for secure communication (for
  message passing as well as file transfer)

\item Replica Catalogue (both LDAP based RC as well as RLS~\cite{RLS})

\end{itemize}

Since GDMP has a much richer set of functionality, additional features
of Globus are used (e.g. Globus IO for client-server
communication).

%
%
\section{\label{implementation}Implementation Details and Comparison}

After some general architectural introduction, we now go more into
detail with the features of GDMP and edg-replica-manager and how the
tools are used in the EDG testbed. For each of the two software tools
we give advantages and disadvantages and thus a critical
discussion. Finally, we compare the two tools directly and point out
for which use case they can be used in a most efficient way.

\subsection{Replica Catalogue Interaction}

Since both our main replication tools GDMP and edg-replica-manager use
a replica catalogue to identify and locate file replicas, we first explain the
replica catalogue interaction and its usage.

In the Globus Toolkit, a simple, centralised replica catalogue is
provided that is based on LDAP technology for storing and retrieving
replica information~\cite{Allcock2001}. In EDG, we developed a wrapper
around the Globus C API and provided a C++ API as well as a simple
command line tool.

On the EDG testbed, this LDAP based Replica Catalogue has been used
but it showed several limitations as outlined in the list
below.  

\begin{itemize}
\item \emph{Performance deterioration with number of entries}:
  Due to the way how the LDAP schema has been chosen for the replica
  catalogue, we experienced low response times (in the order of 30
  seconds to a few minutes) for inserts into the
  Replica Catalogue with the number of entries. If the filenames
  (LFNs) are short (in the order to 10 characters), this problem does not
  occur too often but with long filenames (in the order of 50 to 100
  characters per LFN), there were severe limitations. This is also partly
  due to the overhead of the C++ wrapper.

\item \emph{Centralised, non-scalable}: The LDAP based replica
  catalogue is hosted by a single LDAP server and thus is a single
  point of failure. Based on the previous item, it was identified that
  the catalogue did not easily scale to large amounts of file
  entries. Thus, we needed to impose restrictions on the users to
  limit the amount of inserts within a certain time window.

\item \emph{No high level user command line tool for browsing}: There
  exist a few command line tools provided by Globus and EDG to query
  the catalogue but there are no high level tools for browsing. An
  alternative is a simple Graphical User Interface provided by EDG but
  not deployed on the EDG testbed. Another option is a simple LDAP
  browser.

\item \emph{Schema not flexible}: the LDAP based schema that is
  organised in collections, locations and logical files etc. does not
  allow for a simple extension. 

\item \emph{LFN sub-set of PFN}: there is a severe limitation on file
  naming since the LFN always needs to be a sub-set of the physical file
  name. Thus, all Storage Elements need to have a similar directory
  structure of replicas. This can be a limitation since it imposes
  specific and global configurations of SEs (i.e. all SEs need to be
  configured in a similar way).

\end{itemize}

The based Replica Catalogue tool provided by Globus did not provide
GSI authentication. This was
added by our partners in NorduGrid and then integrated into the
Replica Catalogue server (edg-rc-server) and the Replica Catalogue
API/CLI. However, it was not deployed on the testbed due to the way
the GSI support was added and the low flexibility offered by LDAP in
terms of configuration options. 

EDG and Globus have identified and discussed all issues above and
thus provided a new solution known as Replica Location Service
(RLS)~\cite{RLS}. In later versions of both GDMP and
edg-replica-manager, interfaces to the RLS have been provided and
most of the issues outlined above were eliminated. However, on the EDG
testbed only the LDAP based RC was deployed up to now. RLS will be
part of EDG release 2.0.

\subsection{Mass Storage System Interaction}

Basic file transfer mechanisms like GridFTP allow for secure and
efficient file transfer from one disk server to another. However,
since large amounts of files are not only stored on disks but
also on Mass Storage Systems (MSS) like Castor or HPSS, file replication tools
need to provide a mechanism to transfer files between Storage
Elements, regardless of storage method used (disk
or tape drives managed by a Mass Storage System).

Originally, when we designed and developed GDMP, there was no direct
Grid-enabled interface that allowed for such a file transfer. Thus,
the following solution was applied - primarily to applications in the
High Energy Physics community: a large disk (or a disk pool) is
considered as a first cache and all wide-area transfers are done from
disk to disk. An additional file transfer is then required between the
disk (pool) and the Mass Storage System. Thus, a file replication
step includes a wide-area file transfer as well as a local staging
to/from the Mass Storage System. Such staging interfaces are provided
by GDMP. For further details refer to~\cite{GDMP1}.

In EDG release 1.4, GDMP's interface has been deployed for systems
like Castor and HPSS. 

Obviously, such an additional file copy step can be avoided if the Mass
Storage System provides a direct Grid-enabled interface supporting
security (GSI), GridFTP, virtual organisations and space
management. The Storage Resource Manager (SRM) interface as described
in~\cite{SRM} provides part of that. Several solutions are currently
under development within several projects and are supposed to be
included into EDG release 2.x.

\subsection{GDMP - Grid Data Mirroring Package}

\begin{figure}[t,h]
\centering
\includegraphics[width=80mm]{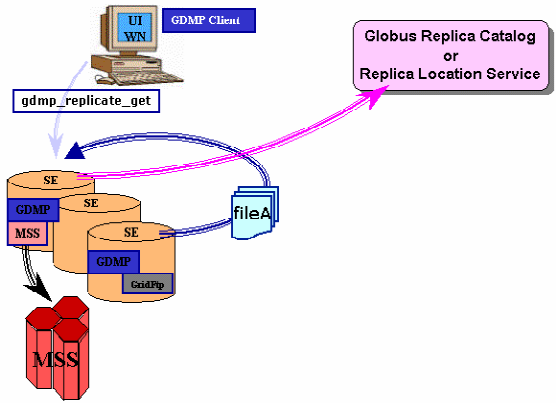}
\caption{File replication/mirroring with GDMP.} \label{figure:GDMP}
\end{figure}

GDMP was a pioneer effort that started initially in the CMS collaboration
(driven by the High Energy Physics community) and it was originally
designed to support file replication in  High Level Trigger
studies. Later, it became a joint project between EDG and PPDG. It
allows for  mirroring of data between Storage Elements through a host
subscription method. The basic interaction is outlined in
Figure~\ref{figure:GDMP}.

GDMP has been enhanced and improved for several years (see
Section~\ref{history}), also based on lots of feedback from application
users in the EDG testbed. Below we list the pros and cons of GDMP
that we and our users experienced in the deployment of GDMP.

\subsubsection{GDMP Advantages}
\begin{itemize}
\item \emph{stable and scalable architecture}: GDMP's
  architecture has proven to be stable and scalable to the needs of
  the basic file replication sites.

\item \emph{Reliable and robust replication}: the transfer mechanisms
  are reliable and robust although we faced a few problems with
  earlier implementations of the GridFTP library.

\item \emph{Retries on error}: if files are not available at the
  time of transfer, the GDMP server takes care of multiple retries and
  thus initiating the file transfer at a later point in time.

\item \emph{File check summing after file transfer}: CRC check summing
  is used to compare the file contents at the beginning and the end of
  a file transfer.

\item \emph{Complex server side logging}: the GDMP server takes care
  of logging all possible events in the file transfer process
  (including staging, subscription etc.). This also allows for
  debugging of file transfers in case of failures.

\item \emph{Users can control file transfer via local catalogues}:
  import, export and local file catalogues can be used to filter files and
  thus reduce the replication process to a specific set of files.

\item \emph{back-ends available for actions to be performed on replication}
  Mass Storage System hooks , automatic replication, post replication
  actions, etc. are provided by the GDMP server.

\item \emph{Mass Storage System interface}: basically for Castor, HPSS
    or equivalent

\end{itemize}

\subsubsection{GDMP Disadvantages}

\begin{itemize}
\item \emph{Designed for site rather than point-to-point replication}: GDMP
  was designed to handle mirroring among sites and not for
  point-to-point replication. Point-to-point replication was another
  requirement that appeared during the usage of GDMP in the EDG
  testbed. In order to respond to this request, the edg-replica-manager has
  been provided.

\item \emph{Several steps involved for replication}: due to the fact that
 GDMP can mirror entire directories with their files based on a
 subscription model, three commands need to executed in order to
 register files in a local catalogue, get them published to remote
 sites and then replicate them. Several users thought that this
 involved too many steps: this has again been addressed in the
 edg-replica-manager at the cost that no subscription is available.

\item \emph{Difficult configuration}: since GDMP has a rather complex
  set of features and offers support for multiple VOs on one server, the
  configuration is rather complex (``difficult''). Some improvements
  could be made as regards the configuration and user authentication
  mechanism.

\item \emph{No space management provided}: space management is beyond
  the scope of GDMP and is the responsibility of the Storage Element
  service (or SRM).

\item \emph{Error messages not always clear}

\item \emph{Errors recovery requires sometimes manual intervention}

\end{itemize}

For more background on GDMP, we refer the reader to~\cite{GDMP1,
  GDMP2}.

\subsection{edg-replica-manager}

The edg-replica-manager~\cite{edg-rm} extends the replica management library in
Globus Toolkit (TM) 2.0 and is a client side tool rather than a
client-server system. It allows for replication and registration 
of files in a Replica Catalogue and works with the LDAP based Globus
Replica Catalog as well as the Replica Location Service (available in
VDT 1.1.7 or higher~\cite{vdt}). In addition, it uses GDMP's staging interface to stage
files to a Mass Storage System. The edg-replica-manager uses the EDG
Replica Catalogue API (in C++).

The edg-replica-manager uses the information service (Globus' MDS is
used in EDG release 1.4) to find out storage locations on a given
Storage Element. It is assumed that basic account management on a
Storage Element is done via tools provided by the GDMP configuration
part. Thus, the edg-replica-manager takes this into account and finds out
where to store files of the particular virtual organisation a user
belongs to. In this way, an end-user only needs to specify the host
name of a given Storage Element and the edg-replica-manager then takes
care of finding the exact source and destination as well as triggering
a staging operation to/from a Mass Storage System.

\begin{figure}[t,h]
\centering
\includegraphics[width=80mm]{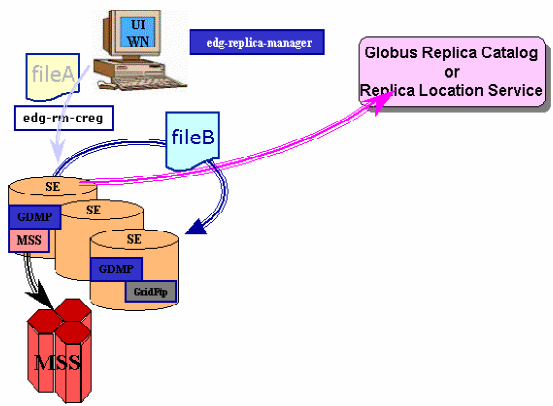}
\caption{edg-replica-manager on the EDG Testbed} \label{figure:edg-rm}
\end{figure}

The basic interaction is outlined in Figure~\ref{figure:edg-rm}. Note
that this tool can be used to transfer files from any of the nodes in
the EDG testbed (i.e. User Interface machine, Worker Node, Storage
Element). A simple command line interface as well as a C++ interface
are provided~\cite{edg-rm}.

Similar to GDMP, we list the pros and cons of the edg-replica-manager
in the following two subsections.

\subsubsection{edg-replica-manager Advantages}
\begin{itemize}
\item \emph{User friendly interface}: since the replica manager has a
  rather small amount of features and initial feedback from our user
  community has been gathered through GDMP, this tool provides a user
  friendly interface.

\item \emph{Functional}: the basic requirement of a replication tool
  is satisfied including that it hides several details of storage
  locations, i.e. detailed storage locations are not required for storing
  and retrieving files - only the LFN is required rather than full PFNs.

\item \emph{Third party transfer available}: using the features of
  GridFTP, a third-party transfer can be triggered from any node where
  the edg-replica-manager client is installed.

\item \emph{GSI authorisation available for Replica Catalogue}: due to
  our modifications to the LDAP based Replica Catalogue server, we also enabled
  GSI authentication for the edg-replica-manager. For RLS, GSI
  authentication is the default option.

\item \emph{Easy configuration}: only a few client side parameters
  need to be set in order to configure the interaction with the
  Replica Catalogue and the Mass Storage System (i.e. GDMP's interface
  to the MSS).
 
\end{itemize}

\subsubsection{edg-replica-manager Disadvantages}
\begin{itemize}
\item \emph{Error messages not always clear}

\item \emph{no roll-back; no transactions}: since edg-replica-manager
  does not have a corresponding server (as it is the case for GDMP),
  no roll-back or transactions are implemented. In addition, there is
  no file checksumming nor centralised logging. In summary, the added
  value that one has with a client-server tool is not gained here.

\item \emph{No complete interface to replica catalogue schema}:
  logical file information like file size or CRC checksumming are not
  supported directly. One needs to use the EDG C++ interface to the
  Replica Catalogue.

\end{itemize}

For more details on the user interface of the edg-replica-manager,
refer to the documentation at~\cite{edg-rm}.

\subsection{Comparison GDMP - edg-replica-manager}

A schematic comparison of the two replication tool is given in
Table~\ref{table:gdmp-vs-rm} and shall assist in choosing which tool
to use for a particular replication requirement.

\begin{table}[h]
\begin{center}
\begin{tabular}{|l|l|}
\hline
{\bf GDMP} & {\bf edg-replica-manager} \\
\hline
replication between SEs only & replication between SEs, \\
                             & UI or CE to SE. \\
Replicates sets of files & replicates single files \\
provides MSS interface & uses GDMP's MSS interface \\
client-server & client side only \\
logical file attributes:  & \\
   (size  times-tamp, etc. & \\ 
  ... extensible) & \\
Subscription model & \\
Event notification & \\
CRC file size check & \\
Support for Objectivity/DB & \\
Automatic retries & \\
Support for multiple VOs & \\
\hline
\end{tabular}
\caption{Comparison: GDMP versus edg-replica-manager}
\label{table:gdmp-vs-rm}
\end{center}
\end{table}

To sum up, the main difference between the ``older'' GDMP and the
``younger'' edg-replica-manager is that the former is a client-sever
tool with a reach set of functionality whereas the later is newer
client side tool only with more stream-lined but smaller set of functionality.

%
%
\section{\label{deployment}Deployment Experience in Several Testbeds}

Our replication software tools have been deployed in various testbeds
as we point out below. The software itself is mainly distributed as part of the
European DataGrid software release, also referred to as \emph{EDG
  release}~\cite{repository}. The EDG release contains all EDG
software, ranging from workload, data, information, fabric and mass
storage management, i.e. it includes all our replication tools as well
as other software. The latest version that has been deployed on the
EDG testbed is EDG release 1.4, our main reference point in the
discussion in this article. 

\subsection{History of Replication Tool Development}
\label{history}

Within the last three years, we gained lots of experience with data
replication tools in a Grid environment. For a complete history of the
development and the basic features that have been included in each
release of the software, we illustrate the replication tool life cycle
in Table~\ref{table:history}. Note that this table also shows when we
stared the edg-replica-manager releases.

\begin{table*}[ht]
\begin{center}
\begin{tabular}{|l|l|}
\hline
{\bf GDMP 1.x} & First prototype of basic SE-SE replication of
Objectivity files \\
September 2000 & Based on Globus 1.1.3\\

\hline
{\bf GDMP 2.x}          & general file replication tools (not only
                          Objectivity files) \\
October 2001            & uses GridFTP + Globus Replica Catalogue\\
                        & full Mass Storage Support \\
\hline
{\bf GDMP 3.x}          & split into client and server side tool \\
April 2002              & improved server functionality/security \\
                        & support for multiple VO \\

\hline
{\bf edg-replica-manager 1.x} & Based on globus-replica-management and
globus-replica-catalog \\
May 2002 & libraries \\

\hline
{\bf edg-replica-manager 2.x} & Several improvements   \\
December 2002 & Replica Location Service (RLS) binding\\

\hline
{\bf GDMP 3.2.x} & RLS + several improvements \\
October 2002 &  \\

\hline
{\bf GDMP 4.0} & Globus 2.2.4 + RH 7.3 gcc 2.95.2 + gcc 3.2 \\
February 2003 & \\
\hline
\end{tabular}
\caption{History of replication tools with their versions and features}
\label{table:history}
\end{center}
\end{table*}

Note that Globus 2.2.x does not support the replica catalog nor the
replica management libraries. Therefore, edg-replica-manager has not
been completely ported to Globus 2.2.4 but we succeeded with GDMP
since there is only the dependency to globus-replica-catalog and EDG
provided a special version of that library.

\subsection{Deployment in Several Testbeds}

Our replication tools were not only used and deployed in the EDG
testbed, but also in a few other environments as we point out
below. Note the GDMP was also part of an early VDT release~\cite{vdt}.

GDMP was first used for High Level Trigger studies (``production'') of HEP 
experiments in 2000/2001 (replication between SEs). In this
environment, we gained our first experience and used the tool in a
``production like'' environment.

Later, GDMP was introduced to the European DataGrid testbed which was
originally set up in autumn 2001. This also resulted in some changes
of user requirements: all user commands needed to be executed from a
User Interface machine or  from Worker Nodes of a Computing Element.
This caused some redesign of the GDMP architecture.

Both tools (GDMP and edg-replica-manager) are used in European and
U.S. testbeds:

\begin{itemize}
\item EDG: ATLAS, CMS, Alice and LHCb stress tests

\item WorldGrid: WorldGrid is the first transatlantic testbed where
  inter-operable between European and U.S. Grid tools has been
  demonstrated~\cite{worldgrid,demos}. 

  As regards the our replication tools: edg-replica-manager was used
  by both CMS and ATLAS applications to move and replicate files
  between U.S. and European sites. GDMP was used as part of the CMS
  MOP environment to replicate set of files produced at several sites.

\item LCG-0: deployed and inter-operable with WorldGrid and GLUE
  testbeds as has been shown in ~\cite{lcg-0}.

\end{itemize}

%
%
\section{\label{future}Conclusion and Future Work}

Within the last three years, we gained lots of experience in
developing and deploying replication tools in a Data Grid
environment. Our first generation tools (GDMP, edg-replica-manager,
API and command line interface replica catalogue) have been successfully
used in some ``production like'' environments as well as in several
testbeds in Europe and in the U.S. All the tools are included in EDG
release 1.4 where they are currently deployed on the EDG application
testbed.

The tools we designed and developed cover client-server as well
as client side tools and thus provide a wide range of possible
design choices. Whereas a client-server tool allows for complex
functionality (including fault tolerance, retries, server side logging,
server side file processing etc.), the configuration is comparably more
complex than for simple client tools like the edg-replica-manager. The
tradeoff in such client-side-only solutions is that many features that
one might want to have for fault tolerance and reliability are
missing. We also gained experience with providing configuration
options to our software tools: in a complex testbed it is of major
importance to keep the configuration as simple as possible. In the
current release, users experience some difficulties with relative complex
configuration options. 

The experience we gained from our first generation tool is used in the
development for the second generation replication tools that will be
provided by EDG in release 2.0. In particular, new services like a
Replica Location Service + Replica Metadata Catalogue, an Optimization
service etc. will be added to the basic functionality of the second
generation tools.

\begin{acknowledgments}

Special thanks to Asad Samar (one of the co-founders of GDMP) and
Aleksandr Konstantinov who participated to the code development of
GDMP. We also thank our user community and our colleagues for valuable
feedback and input.

This work was partially funded by the European Commission program
IST-2000-25182 through the EU DataGrid Project.
\end{acknowledgments}


\end{document}